\documentclass[twocolumn,aps,showpacs,eqsecnum,10pt]{revtex4} 
\usepackage{latexsym}
\usepackage{amsthm}
\usepackage{amsmath}
\usepackage{amssymb}
\usepackage{epsfig}

\newtheorem{proposition}{Proposition}
\newtheorem{theorem}{Theorem}

\theoremstyle{definition}

\newcommand{\bra}[1]{\langle #1|}
\newcommand{\ket}[1]{| #1 \rangle }
\newcommand{\tr}[1]{{\rm tr}[#1]}

\newcommand{\be}{\begin{eqnarray}}
\newcommand{\ee}{\end{eqnarray}}

\newcommand{\one}{{\openone}}

\newcommand{\cE}{{\cal E}}
\newcommand{\cI}{{\cal I}}

\newcommand{\cT}{{\cal T}}
\newcommand{\cS}{{\cal S}}
\newcommand{\cH}{{\cal H}}
\newcommand{\cA}{{\cal A}}

\begin{document}
\title{Unambiguous comparison of unitary channels}
\author{Michal Sedl\'ak$^{1,3}$,
M\'ario Ziman$^{1,2}$}
\affiliation{
$^{1}$Research Center for Quantum Information, Institute of Physics, Slovak Academy of Sciences, D\'ubravsk\'a cesta 9, 845 11 Bratislava,  Slovak Republic \\
$^{2}$Faculty of Informatics, Masaryk University, Botanick\'a 68a, 602 00 Brno, Czech Republic\\
$^{3}${\em Quniverse}, L{\'\i}\v{s}\v{c}ie \'{u}dolie 116, 841 04 Bratislava, Slovakia}
\begin{abstract}
We address the problem of unambiguous comparison of a pair of unknown
qudit unitary channels. Using the framework of process positive operator
valued measures
(PPOVM) we characterize all solutions and identify the optimal ones.
We prove that the entanglement is the key ingredient in designing the optimal experiment for comparison of unitary channels. Without entanglement the optimality can not be achieved. The proposed scheme is also experimentally feasible.
\end{abstract}

\maketitle

\section{Introduction}
The unavoidable uncertainty of quantum predictions represents one of
the key features of quantum theory \cite{helstrom,peres_book}.
It might seem surprising at first sight,
but even in probabilistic theories there are problems in which error-free
nontrivial conclusions can be based on single experimental events (clicks).
Consider, for example,
the Stern-Gerlach experiment in which the spin of a particle is measured
along the $z$ axis. Finding the outcome {\it spin up} implies that the
particle was not for sure in the {\it spin down} state. In general,
this information does not seem to be very useful. However,
if we have additional information that the spin was either
$\varrho$, or {\it spin down}, then the outcome {\it spin up} identifies
the spin state $\varrho$.

A class of problems extending this example is known as unambiguous
identification problems. Over last decades authors have investigated
unambiguous discrimination of states
\cite{ivanovic,dieks,peres,jaeger,kleinmann,hayashi,zhang},
channels \cite{wang,chefles} and observables \cite{teiko}.
All these works are showing that single clicks can give us nontrivial
information about all types of quantum devices. In their seminal work
Barnett et al. \cite{barnett} introduced the concept of unambiguous state
comparison. In this problem an experimentalist is given two
preparators, each producing a single quantum system in a particular
pure state. The aim is to compare produced states. It turns out that in all
possible experiments only the difference of the compared states
can be unambiguously concluded. We can never confirm experimentally (without
making some error) that two unknown pure states are the same.
Different versions of unambiguous state comparison have been investigated
in \cite{chefles2,kleinman2,sedlak}.

The goal of this paper is to investigate a comparison of quantum
devices implementing unknown unitary channels. Such universal
comparator of unitary channels can be of use, for instance, in
the calibration and testing of the quality of elementary quantum gates.

Quantum channels are tested in two steps. First we prepare a so-called
test state and apply the channel. After that the output state is
measured. Therefore, it is natural to employ a state comparator
to compare the channels. As we shall see these two problems
are indeed closely related, but there are also important differences
concerning the optimal strategies. We shall elaborate on this point later.

The paper is organized as follows: in Section II the problem is reformulated
in the framework of process positive operator valued measures (PPOVM) and existence of a solution is shown. The optimal
solution is described in Section III together with its uniqueness. The last
section is left for summary and conclusions.

\section{Formulation of the problem}

Consider we are given two black boxes implementing unknown unitary
channels $\cE_U$ and $\cE_V$ on qudit, i.e. $d$-dimensional
quantum system. Our task is to unambiguously decide
whether the black boxes perform the same unitary channels, or not.
More formally, whether a process implemented on $D=d\times d$ dimensional quantum system by the pair of devices
is described by a channel $\cE_U\otimes\cE_V$ with $U\neq V$,
or by a channel $\cE_U\otimes\cE_U$.
As in any comparison
problem we implicitly assume that the probability that the channels
 are the same is nonzero. Otherwise the problem would be senseless.

Let us note that unlike preparators (represented by states)
the processes (associated with channels) can be used sequentially.
In general, this is an important difference between the usage of
preparators and processes providing us with a resource of a potential use.
However, it does not give us any advantage in the case of
the considered comparison problem. In particular, one cannot
distinguish whether the product
of two unknown unitary channels is $\cE_U\circ\cE_V$ (for $U\neq V$),
or $\cE_U\circ\cE_U$, because for any unitary operator $W$ there exist unitary
operators $U,V\neq W$ such that $W^2=UV$.

\begin{figure}
\begin{center}
\includegraphics[width=4.5cm]{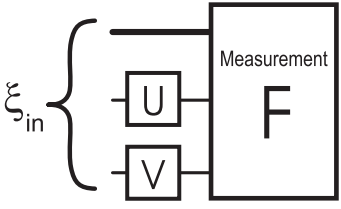}
\caption{Experiment for comparison of two unitary channels $\cE_U, \cE_V$.}
\label{exp1}
\end{center}
\end{figure}

The experimental procedure for the
comparison is illustrated in Figure \ref{exp1}. Using each of the quantum
boxes at most once the experiment will end by a measurement, whose outcome
uniquely determines our conclusion. In particular, the experiment consists
of three steps. At first, we prepare a so-called test
state $\xi\in\cS(\cH_{\rm anc}\otimes\cH_d\otimes\cH_d)$,
where $\cH_{\rm anc}$ is the Hilbert space of some ancilliary system.
After that black boxes are applied and a measurement $F$
on the whole system including the ancilla is performed. Measurement
outcomes are associated with effects
$F_{\rm same},F_{\rm diff},F_{?}$ forming a three-valued POVM, i.e.
\be
\nonumber
O\leq F_{\rm same},F_{\rm diff},F_?\leq I\,;
\qquad F_{?}+F_{\rm same}+F_{\rm diff}=I
\, .
\ee
As in any unambiguous identification problem the inconclusive outcome
$F_?$ is needed in order to make the conclusive outcomes
$F_{\rm same},F_{\rm diff}$ unambiguous. In fact, we shall see explicitly that
$F_{?}\neq O$.
An outcome $x\in\{{\rm same,\ diff,\ ?}\}$
is observed with the probability
\be
p_x(U\otimes V)=\tr{F_x(\cI_{\rm anc}\otimes\cE_U\otimes\cE_V)[\xi]}\,,
\ee
where $\cE_U[\cdot]=U\cdot U^\dagger,\cE_V[\cdot]=V\cdot V^\dagger$ are
unitary channels implemented by the black boxes.

Our goal is to characterize all possible experiments (determined
by pairs $\xi,F$) performing the unambiguous comparison of
unitary channels and identify the optimal strategy. The figure of merit
for the optimization will be specified in details later. The analysis
can be significantly simplified by adopting a framework introduced
in Refs. \cite{ziman1,dariano}. According to work \cite{ziman1} each experiment
measuring some parameters of a channel can be decribed by the so-called
process POVM (PPOVM). The key idea behind the PPOVM framework is that each
experiment can be understood as a fictitious experiment using the maximally
entangled state as the test state. In particular, in this framework
channels acting on $D$ dimensional quantum systems are represented via Choi-Jamiolkowski isomorphism \cite{choi,jamilk}
by operators
\be
\omega_\cE=(\cI\otimes\cE)[\Psi^+_D]
\ee
defined on $D^2$-dimensional Hilbert space,
where $\Psi_D^+=\ket{\Psi_D^+}\bra{\Psi_D^+}$ and
$\ket{\Psi_D^+}=\sum_{j=1}^D \ket{j}\otimes\ket{j}$. Let us note
that in this parametrization $\tr{\omega_\cE}=D$. For a given
pair of test state $\xi$ and observed effect $F_x$ there
exists an operator $O\leq M_x\leq I_D\otimes I_D$ on $D^2$-dimensional
Hilbert space such that
$p_x=\tr{\omega_\cE M_x}=\tr{(\cI\otimes\cE)[\xi]F_x}$
for all channels $\cE$. Moreover, the
normalization condition $\sum_x M_x=\varrho^T\otimes I_D$ holds,
where $\varrho$ is a state
of $D$-dimensional system and
$\varrho^T$ denotes the matrix transposition with respect to basis
used in the definition of the maximally entangled state $\Psi_D^+$.
Let us note that for a given PPOVM, i.e.
a set of positive operators $M_1,\dots,M_n$ such that
$\sum_x M_x=\varrho^T\otimes I_D$ there exists many different
experiments with different choices of test states and POVMs \cite{ziman1}.
In particular, consider a PPOVM such that
$M_j=\varrho^T\otimes F_j$ for all $j$. Since the identity
\be
\nonumber
\tr{\cE[\varrho]F_j}=
\tr{(\cI\otimes\cE)[\Psi^+_D] (\varrho^T\otimes F_j)}=
\tr{\omega_\cE M_j}
\ee
holds for all qudit channels $\cE$ and all qudit operators
$\varrho,F$, it follows that this type of PPOVM can be realized by using
a single ancilla-free test state $\varrho$ and performing the
measurement described by POVM consisting of positive operators $F_j$.

\subsection{Requirements on unambiguous comparators}

Translating the comparison problem into PPOVM framework we set $D=d^2$
and associate the two black boxes acting on $d$-dimensional systems
with operators
\be
\omega_{U\otimes U}&=& (I_D\otimes U\otimes U)
\Psi_D^+ (I_D\otimes U^\dagger\otimes U^\dagger)\, ,\\
\omega_{U\otimes V}&=& (I_D\otimes U\otimes V)
\Psi_D^+(I_D\otimes U^\dagger\otimes V^\dagger)\, ,
\ee
where $\Psi_D^+=\ket{\Psi_D^+}\bra{\Psi_D^+}$ and
$\ket{\Psi_D^+}_{1234}=\ket{\Psi_d^+}_{13}\otimes\ket{\Psi_d^+}_{24}\in
\cH_d^{\otimes 4}$.
Operators $M_{\rm same},M_{\rm diff},M_{\rm ?}$ defining
the PPOVM have to satisfy following no-error conditions ensuring
the unambiguity of the corresponding conclusions:
\be
\nonumber
p_{\rm diff}(U\otimes U)&=&\tr{\omega_{U\otimes U}M_{\rm diff}}=0\\
\nonumber
p_{\rm same}(U\otimes V)&=&\tr{\omega_{U\otimes V}M_{\rm same}}=0
\ee
for all $U,V\in U(d)$, where $U(d)$ denotes the group of unitary operators
on $d$-dimensional Hilbert space.

Defining average channels as
\be
\cA[X]&=&\int_{U(d)}dU U X U^\dagger \,,\\
\cT[Y]&=&\int_{U(d)}dU (U\otimes U) Y (U^\dagger\otimes U^\dagger)\,,
\ee
the above conditions can be equivalently rewritten as
\be
\label{eq:diff_1}
0&=&\tr{(\cI_{12}\otimes\cT_{34})[\Psi_D^+] M_{\rm diff}}\,,\\
\label{eq:same_1}
0&=&\tr{(\cI_{12}\otimes\cA_3\otimes\cA_4)[\Psi_D^+] M_{\rm same}}\, ,
\ee
because all the relevant operators are positive.
The actions of the twirling channel $\cT$ and the average channel $\cA$
are derived in Appendices. In particular,
\be
\cA[X]&=&\tr{X}\frac{1}{d}I_d\,, \\
\cT[Y]&=&\frac{\tr{YP_+}}{d_+}P_++\frac{\tr{YP_-}}{d_-}P_-\, ,
\ee
where $P_\pm$ are projectors onto symmetric and antisymmetric
subspaces of $\cH_d\otimes\cH_d$, respectively, and
$d_\pm=\tr{P_\pm}=d(d\pm 1)/2$ are the corresponding dimensions
of these subspaces. Let us note that $P_\pm=\frac{1}{2}(I\pm S)$, where
$S$ is the swap operator acting as $S\ket{\psi}\otimes\ket{\varphi}=
\ket{\varphi}\otimes\ket{\psi}$ for all $\psi,\varphi\in\cH_d$.
Using these expressions we obtain
\be
(\cI_{12}\otimes\cA_3\otimes\cA_4)[\Psi_D^+]=\frac{1}{d^2}I_d^{\otimes 4}
\ee
and since
\be
\nonumber
\cT[\ket{jm}\bra{kn}]&=&
\sum_{s=\pm}\frac{1}{d_s}\bra{kn}P_s\ket{jm}P_s
\\\nonumber
&=&\sum_{s=\pm} \frac{1}{2d_s}(\delta_{jk}\delta_{mn}+s\delta_{jn}\delta_{mk})P_s
\ee
we have
\be
\nonumber
\omega_\cT&=&(\cI_{12}\otimes\cT_{34})[(\Psi_d^+)_{13}\otimes(\Psi_d^+)_{24}]
\\ \nonumber
&=& \sum_{j,k,m,n} \ket{jm}_{12}\bra{kn}\otimes \cT_{34}[\ket{jm}_{34}\bra{kn}]
\\\nonumber
&=&
\frac{1}{2d_+}\sum_{j,m}\Big[\ket{jm}\bra{jm}+\ket{jm}\bra{mj}\Big]\otimes P_+
\\\nonumber
& & +\frac{1}{2d_-}\sum_{j,m}[\ket{jm}\bra{jm}-\ket{jm}\bra{mj}]\otimes P_-
\\\nonumber
&=&
\frac{1}{4d_+}\sum_{j,m}[(\ket{jm}+\ket{mj})(\bra{jm}+\bra{mj})]\otimes P_+
\\\nonumber
& & +\frac{1}{4d_-}\sum_{j,m}[(\ket{jm}-\ket{mj})(\bra{jm}-\bra{mj})]\otimes P_-
\\\nonumber
&=& \frac{1}{d_+}P_+\otimes P_++\frac{1}{d_-}P_-\otimes P_-\,.
\ee
Putting all formulas together the
conditions in Eqs.(\ref{eq:diff_1}),(\ref{eq:same_1})
take the form
\be
0&=&\tr{\omega_\tau M_{\rm diff}}\,,\\
0&=&\frac{1}{d^2}\tr{I_d^{\otimes 4} M_{\rm same}}=\tr{M_{\rm same}}\,.
\ee
Since $M_{\rm same},M_{\rm diff}$ are positive operators it follows
that $M_{\rm same}=O$ and $M_{\rm diff}$ has support in the orthocomplement of $\omega_\tau$.
Consequently, we can unambiguously conclude only
that the unitary channels are different. We can formulate the following
proposition.
\begin{proposition}
If a PPOVM $M_{\rm same},M_{\rm diff},M_?$ describes an unambiguous comparison
of arbitrary unitary channels, then necessarily
\be
\nonumber
&{\rm supp}M_{\rm diff}\perp{\rm supp}\omega_\cT
\,;\qquad
M_{\rm same}=O\,;& \\
&M_{?}=\varrho^T\otimes I_D-M_{\rm diff}\, ,&
\ee
for some state $\varrho\in\cS(\cH_d\otimes\cH_d)$.
\end{proposition}

\section{Optimal unambiguous comparator}

Following the previous section as a figure
of merit for unambiguous comparators of unitary
channels we shall use the average conditioned probability of revealing
their difference
\be
\nonumber
\overline{p}_{\rm diff}&=&\int_{U(d)\times U(d)} dUdV\,p_{\rm diff}(U\otimes V)\\
\nonumber
&=& \tr{(\cI_{12}\otimes\cA_{3}\otimes\cA_{4})[\Psi^+_{D}] M_{\rm diff}}\\
&=&
\frac{1}{d^2}\tr{M_{\rm diff}}\, .
\label{psucc}
\ee
The overall average success probability equals $(1-\eta_{\rm same})
\overline{p}_{\rm diff}$, where $\eta_{\rm same}\neq 0$ is the prior
probability for channels being the same. This prior is independent
of the particular PPOVM $\{M_{\rm diff},M_? \}$ and therefore
we shall use only the conditional average probability to
evaluate the quality of the unambiguous comparison strategy. Our task is to maximize the conditional success probability
$\overline{p}_{\rm success}\equiv\overline{p}_{\rm diff}$ by finding
a positive operator $M_{\rm diff}$ defined on $\cH_D\otimes\cH_D$
together with a state $\varrho\in\cS(\cH_D)$ such that also the operator
$M_{?}=\varrho^T\otimes I_D-M_{\rm diff}$ is positive.
Before specifying the optimal solution let us prove the following
upper bound on the success probability.

\begin{theorem}
If a process POVM consisting of positive operators
$M_{\rm diff},M_{?}$ with normalization
$M_{\rm diff}+M_{?}=\varrho^T\otimes I_D$ unambiguously
compares an arbitrary pair of unitary channels, then
\be
\overline{p}_{\rm success}\leq \frac{d+1}{2d}\, .
\label{psucnerov}
\ee
\end{theorem}
\begin{proof}
The validity of the no-error condition $\tr{\omega_\cT M_{\rm diff}}=0$
implies that supports of $M_{\rm diff}$ and $\omega_\cT$ are orthogonal.
Let us denote by $\ket{s_1},\dots,\ket{s_{d_+}}$,
$\ket{a_1},\dots,\ket{a_{d_-}}$ the vectors forming orthonormal bases of
symmetric and antisymmetric subspaces of $\cH_d\otimes\cH_d$, respectively.
Then ${\rm supp}\, \omega_\cT={\rm span} \{\ket{s_j\otimes s_k},\ket{a_m\otimes a_n}\}$,
where $j,k=1,\dots,d_+$ and $m,n=1,\dots,d_-$, and because
of the mentioned orthogonality
\be
\label{eq:supp_diff}
{\rm supp}M_{\rm diff}\subset{\rm span}\{\ket{s_j\otimes a_n},\ket{a_n\otimes s_j}\}\, .
\ee
It follows that in a spectral form
\be
M_{\rm diff}=\sum_\alpha \lambda_\alpha \ket{\phi_\alpha}\bra{\phi_\alpha}\, ,
\label{spektr1}
\ee
where $0\le\lambda_\alpha\le 1$ and
\be
\ket{\phi_\alpha}=\sum_{nj} c_{nj}^\alpha\ket{a_n\otimes s_j}+
d_{jn}^\alpha\ket{s_j\otimes a_n}\, .
\label{vlvekt1}
\ee
Consequently,
\be
M_{\rm diff}=\sum_n \ket{a_n}\bra{a_n}\otimes A_n+
\sum_n B_n\otimes\ket{a_n}\bra{a_n}+R\,, \nonumber
\ee
with
\be
\nonumber
A_n&=&\sum_\alpha\lambda_\alpha\sum_{jl} c_{nj}^\alpha \overline{c_{nl}^\alpha}
\ket{s_j}\bra{s_l}\,;\\
\nonumber
B_n&=&\sum_\alpha\lambda_\alpha\sum_{jl} d_{nj}^\alpha \overline{d_{nl}^\alpha}
\ket{s_j}\bra{s_l}\,;\\
\nonumber
R&=&
\sum_\alpha\lambda_\alpha\left[
\sum_{m\neq n, j,l} c_{mj}^\alpha\;\overline{c_{nl}^\alpha} \ket{a_m\otimes s_j}
\bra{a_n\otimes s_l}+\right.\\ \nonumber
& & \qquad\quad
+\sum_{m\neq n, j,l} d_{jm}^\alpha\;\overline{d_{ln}^\alpha} \ket{s_j\otimes a_m}
\bra{s_l\otimes a_n}+\\ \nonumber
& & \qquad\quad
+\sum_{m,n, j,l} c_{mj}^\alpha\;\overline{d_{ln}^\alpha} \ket{a_m\otimes s_j}
\bra{s_l\otimes a_n}+\\ \nonumber
& & \qquad\quad
+\left.\sum_{m,n, j,l} d_{jm}^\alpha\;\overline{c_{nl}^\alpha} \ket{s_j\otimes a_m}
\bra{a_n\otimes s_l}
\right]\, .
\ee
Since $\tr{R}=0$ we get for the average success probability
\be
\overline{p}_{\rm success}=\frac{1}{d^2}\sum_{n=1}^{d_-}(\tr{A_n}+\tr{B_n})\,.
\ee
The operators $A_n,B_n$ have the form of positive sum of one-dimensional
projectors, hence they are positive.

Let us evaluate the mean value of operator $M_{?}=
\varrho^T\otimes I-M_{\rm diff}$ in a pure state associated
with the vector $\ket{s_j\otimes a_n}$. Due to the required positivity
of $M_{?}$ we get the inequality
\be
0\leq \bra{s_j\otimes a_n}M_{?}\ket{s_j\otimes a_n}=
\bra{s_j}\varrho^T-B_n\ket{s_j}\, .
\ee
Similarly, also the inequality
\be
\nonumber
0&\leq& \bra{a_n\otimes s_j}M_{?}\ket{a_n\otimes s_j}
\\
&\leq&\bra{a_n}\varrho^T\ket{a_n}-\bra{s_j}A_n\ket{s_j}\,
\label{nerov0}
\ee
holds. These two inequalities can be used to bound the
trace of the density operator $\varrho^T$ as follows
\be
\nonumber
\tr{\varrho^T}&=&\sum_n \bra{a_n}\varrho^T\ket{a_n}+\sum_j
\bra{s_j}\varrho^T\ket{s_j}\\
\nonumber
&\ge&\sum_n \bra{s_k}A_n\ket{s_k} +\sum_j \bra{s_j}B_m\ket{s_j}\\
&\ge& \bra{s_k}\sum_n A_n\ket{s_k}+\tr{B_m}\,,
\label{nerov1}
\ee
where we used the fact that by definition operators $B_m$ have
support only on the symmetric subspace. The inequality holds
for all choices of $k$ and $m$. Moreover, since $\tr{\varrho^T}=1$
and $B_m$ is positive, i.e. $\tr{B_m}\ge 0$, we obtain that also
\be
\bra{s_k}\sum_n A_n\ket{s_k}\leq 1\, .
\label{nerov2}
\ee
for all $k$. Using these inequalities the success probability
can be upper bounded as follows
\be
\nonumber
\overline{p}_{\rm success}&=&\frac{1}{d^2}
\left(
\sum_{j=1}^{d_+} \bra{s_j}\sum_{n=1}^{d_-}A_n\ket{s_j}+\sum_{m=1}^{d_-}\tr{B_m}
\right)\\\nonumber
&=&\frac{1}{d^2}\left[
\sum_{m=1}^{d_-}\left(\bra{s_m}\sum_{n=1}^{d_-}A_n\ket{s_m}+\tr{B_m}
\right)
+\right.\\ \nonumber
& &\qquad\quad
+\left.\sum_{j=d_-  +1}^{d_+} \bra{s_j}\sum_{n=1}^{d_-}A_n\ket{s_j}\right]\\
&\le& \frac{1}{d^2}(d_-+d)=\frac{d_+}{d^2}=\frac{d+1}{2d}\, ,
\ee
which proves the theorem.
\end{proof}

\subsection{Antisymmetric test states}
In what follows we shall design a process POVM saturating
the upper bound on the success probability. In particular, for operators
\be
M_{\rm diff}=\xi^T\otimes P_+\;,\qquad M_?=\xi^T\otimes P_-\, .
\label{asymtest}
\ee
the success probability equals
\be
\overline{p}_{\rm success}=\frac{1}{d^2}\tr{M_{\rm diff}}=
\frac{1}{d^2}\tr{\xi^T\otimes P_+}=\frac{d_+}{d^2}\, ,
\ee
hence the upper bound is saturated. Let us note that the state
$\xi$ is not arbitrary, because the support of $M_{\rm diff}$
must be orthogonal to support of $\omega_\cT$ (see Eq.(\ref{eq:supp_diff})).
It implies that the state $\xi$ has support only on antisymmetric
subspace. We shall call such states antisymmetric. Similarly, if the support of a state is only in symmetric subspace we denote it as symmetric state.

The form of PPOVM in Eq. (\ref{asymtest}) suggests that one possible experimental
realization consists of the folowing steps: i) prepare a two-qudit antisymmetric state $\xi$; ii) insert each qudit
into different black box; iii) measure a two-valued observable described by POVM $F_{\rm diff}=P_{+}$ and $F_{?}=P_{-}$, which identifies
the exchange symmetry of the joint state of the two-qudit system.

The test state $\xi$ is antisymmetric. If $U=V$ the action of the apparatuses
preserves the symmetry, i.e. the output state remains antisymmetric
and in such case $F_{?}$ must be observed. For $U\neq V$
the measurement outcome cannot be predicted with certainty, so both outcomes $F_{\rm diff}, F_{?}$
have nonvanishing probability of occurence. However, if
an outcome $F_{\rm{diff}}$ is observed, we can unambiguously conclude that $U$ and $V$
are different.

\subsection{Symmetric test states}
Alternatively, we can consider a process POVM
\be
M_{\rm diff}=\xi^T\otimes P_-\;,\qquad M_?=\xi^T\otimes P_+\,
\ee
satisfying all the constraints providing
$\xi$ has support in the symmetric subspace. For this choice the
success probability reads
\be
\label{eq:sym_succ}
\overline{p}_{\rm success}=\tr{\xi^T\otimes P_-}=\frac{d_-}{d^2}=\frac{d-1}{2d}
\,,
\ee
which is not optimal. Such PPOVM describes an experiment in which
a "symmetric`` test state is used.
The same measurement is carried out as in the antisymmetric case,
but the role of conclusive
and inconclusive results is exchanged, i.e. $F_{\rm diff}=P_-$
and $F_?=P_+$.

As we have mentioned at the beginning
of this paper one possibility how to tackle
the problem of unambiguous comparison of unitary channels is to adopt
the universal comparison machines for states. Consider a pair of unitary channels applied on independent systems
initially prepared in the same state. If $U=V$, then the resulting states
are still described by the same state. However, if $U\neq V$, then the
output states can be different. That is the state comparator can be used
to find out whether the output states are different, which means that
the unitary channels are different as well. In the language of
channel comparison the described strategy can be interpreted as
a strategy with a symmetric factorized test state
$\xi=\ket{\varphi\otimes\varphi}\bra{\varphi\otimes\varphi}$.
Since the optimal state comparison is based on projective measurement
described by projectors $P_\pm$, the value of the success probability
is given in Eq.(\ref{eq:sym_succ}).

\subsection{Uniqueness of optimal solution}
In previous paragraphs we have shown that optimal strategy for comparison
of unitary channels saturates the upper bound on probability of success
imposed by Theorem 1. It means that PPOVM elements of each optimal strategy
have to saturate all inequalities used in proof of this theorem. Analyzing
this fact we can characterize all optimal strategies.
\begin{theorem}
If a process POVM $\{ M_{\rm diff},M_{?}\}$ with normalization
$\varrho^T\otimes I_D$ unambiguously compares arbitrary pair of unitary channels with $\overline{p}_{\rm success}=\frac{d+1}{2d}$, then
\be
M_{\rm diff}=\varrho^T\otimes P_+\;,\qquad M_?=\varrho^T\otimes P_-\, ,
\ee
where $\varrho$ is a state with a support belonging only to the antisymmetric
subspace of $\cH_d\otimes\cH_d$.
\end{theorem}
\begin{proof}
Saturation of inequality (\ref{nerov2}) for $k=d_+$ together with
inequality (\ref{nerov1}) implies that $\tr{B_n}=0$ for all $n$. Consequently,
positivity of operators $B_n$ implies $B_n=0$ for all $n$ i.e. coefficients
$d_{jn}^\alpha$ vanish. This in turn requires
\be
\bra{s_k}\sum_n A_n\ket{s_k}=1
\label{nerov3}
\ee
for all $k$. Using Eq. (\ref{nerov0}) and Eq. (\ref{nerov3}) we get
\be
\nonumber
1=\sum_n \bra{s_k}A_n\ket{s_k}\leq \sum_n \bra{a_n}\varrho^T\ket{a_n}\leq
\tr{\varrho^T}=1\,,
\ee
thus, $\sum_n \bra{a_n}\varrho^T\ket{a_n}=1$. Due to positivity of
$\varrho^T$ we obtain $\bra{s_j}\varrho^T\ket{s_j}=0$ for all $j$. This
tells us that $\varrho^T$ has support only on antisymmetric states.
Since the used transposition is defined with respect to a product basis,
the antisymmetric states preserve their antisymmetry, i.e. the
state $\varrho$ is antisymmetric as it is stated in the theorem.

Using the spectral form (\ref{spektr1}) and Eq. (\ref{vlvekt1}) we can rewrite $M_{\rm diff}$ as:
\be
M_{\rm diff}=\sum_j C_j\otimes\ket{s_j}\bra{s_j}+H\,,
\ee
with
\be
\nonumber
C_j&=&\sum_\alpha\lambda_\alpha\sum_{nm} c_{nj}^\alpha \overline{c_{mj}^\alpha}
\ket{a_n}\bra{a_m}\,;\\
\nonumber
H&=& \sum_\alpha\lambda_\alpha
\sum_{j\neq l, m,n} c_{mj}^\alpha\;\overline{c_{nl}^\alpha} \ket{a_m\otimes s_j}
\bra{a_n\otimes s_l}
\ee
We rewrite also the probability of success [Eq. (\ref{psucc})]
in terms of $C_j$ and because the operator $H$ is traceless we get
\be
\overline{p}_{\rm success}=\frac{1}{d^2}\sum_j \tr{C_j}.
\ee
Positivity of $M_{?}=\varrho^T\otimes\one-
\sum_j C_j\otimes\ket{s_j}\bra{s_j}-H$ implies
\be
0\leq \bra{a\otimes s_j}M_{?}\ket{a\otimes s_j}=
\bra{a}\varrho^T-C_j\ket{a}\, ,
\ee
where $\ket{a}$ is arbitrary vector from $\cH_D$. Hence,
we have that
operator $\varrho^T-C_j$ is positive for all $j$ and consequently that
$1=\tr{\varrho^T}\geq \tr{C_j}$. Saturation of inequality (\ref{psucnerov})
requires $\tr{C_j}=1$ for all $j$, which in turn implies
$\tr{\varrho^T-C_j}=0$. This together with the positivity of operator
$\varrho^T-C_j$ enables us to conclude that $C_j=\varrho^T$ for all $j$.
The operators $M_{\rm diff}$,$M_{\rm ?}$ therefore read
\be
M_{\rm diff}=\varrho^T\otimes P_+ +H\,, \nonumber\\
M_{\rm ?}=\varrho^T\otimes P_- -H\,. \nonumber
\ee
The support of the selfadjoint operator
$H$ is orthogonal to the support of the operator
$\varrho^T\otimes P_-$. Since $H$ is traceless it has both positive and
negative eigenvalues unless $H=O$. However, positive eigenvalues of
$H$ would spoil positivity of $M_{\rm ?}$, so the operator $H$ must vanish,
which concludes the proof.
\end{proof}

\section{Conclusions}
The goal of this paper was to find an optimal strategy
for comparison of two unknown unitary channels. Exploiting the
framework of process POVM we have shown that the optimal
strategy achieves the average conditional success probability
$\overline{p}_{\rm success}=(d+1)/(2d)$. An interesting observation
is that the optimal strategy for comparison of unitary channels
is very closely related to the comparison of pure states. In fact, the
optimal state comparison is based on the implementation of the two-valued
projective measurement measuring the exchange symmetry of the bipartite
states. Outcomes are associated with projectors $P_\pm$ onto symmetric and antisymmetric
subspaces of the joint Hilbert space.
The optimal procedure for the comparison of unitary channels is
exploiting the same measurement, but the outcomes are interpreted
in the opposite way. Whereas for comparison of pure states the projector
$P_+$ corresponds to the inconclusive result, for unitaries
this projector is associated with the unambiguous conclusion
that the channels are different. Similarly, the projector $P_-$
indicates the difference of compared pure states, but corresponds
to no conclusion for unitaries. In both cases, the unambiguous
conclusion that the states, or unitaries are the same, cannot
be made.

Devices implementing quantum channels are tested indirectly
via their action on quantum states. In the experiment the unknown
apparatuses are probed by some test states. We have shown that
the optimal solution is achieved if and only if the test state
is antisymmetric, i.e. its support is only in antisymmetric subspace.
Let us note that if a state is separable, then necessarily
its support contains product vectors. However, by definition there is no
antisymmetric product vector, hence the support of
each antisymmetric state does not contain any product vectors.
Consequently, each antisymmetric state is necessarily entangled.
In conclusion, the entanglement is the key ingredient for comparison
of unitary channels. It enhances the success probability to reach the
optimal value.

Let us note that the proposed optimal strategy is feasible in current
quantum information experiments with photons and ions. In particular,
in the qubit version the experiment consists of preparation of
a singlet, application of the unknown single-qubit unitary channels
on individual qubits and a projective measurement consisting of the
projection onto a singlet, or arbitrary other maximally entangled state.
As the measurement we can use, for instance, the Bell measurement,
but it is not necessary. Moreover, for the comparison of few qubit
unitary channels mixed test states are allowed.

\section*{ACKNOWLEDGMENTS}
This work was supported by the European Union project
QAP 2004-IST-FETPI-15848 and by APVV via projects
RPEU-0014-06 and APVV-0673-07 QIAM. Authors would like
to thank Vladim\'\i r Bu\v zek and Daniel Reitzner for
reading the manuscript.

\appendix


\section{Average unitary channel}
\label{cdelta}
In this section we shall prove that the action of the
average unitary channel can be expressed as
\be
\label{eq:last}
\cA[X]=\int_{U(d)}dU\, UXU^\dagger = \frac{\tr{X}}{d}I\, ,
\ee
where $dU$ is the unique {\it Haar invariant measure} defined on the
group of unitary operators $U(d)$. By definition the image $\cA[X]$ of any
operator $X$ must commute with all unitary operators, i.e.
$[\cA[X],U]=0$ for all $U\in U(d)$. The Schurr lemma implies that
$\cA[X]=c(X)I$. The transformation $\cA$ is by definition trace-preserving.
That is, $\tr{X}=c(X)\tr{I}= c(X)d$. It follows that $c(X)=\tr{X}/d$,
hence the Eq.(\ref{eq:last}) holds.

\section{Twirling channel}
We shall prove that the action of the twirling channel
\be
\label{eq:twirling_form_1}
\cT[X]=\int_{U(d)} dU\, U\otimes UX U^\dagger\otimes U^\dagger\,,
\ee
on
selfadjoint operators $X$ takes the form
\be
\label{eq:twirling}
\cT[X]=\frac{\tr{XP_+}}{d_+}P_++\frac{\tr{XP_-}}{d_-}P_-\, .
\ee

The properties of Haar invariant measure $dU$ implies that
the operator $\cT[X]$ commutes with all unitary operators of the type
$U\otimes U$. If $X$ is selfadjoint, then $\cT[X]$ is also selfadjoint
and $\cT[X]=\sum_j x_j P_j$, where $x_j$ are real eigenvalues
and $P_j$ are the corresponding eigenprojectors.
The commutation of $\cT[X]$ with unitaries $U\otimes U$ implies
that $[P_j,U\otimes U]=0$ for all $U$. The subspaces
$\cH_j=P_j(\cH_d\otimes\cH_d)
=\{\psi\in\cH_d\otimes\cH_d\ {\rm such\ that}\ P_j\psi=\psi\}$ are invariant
under the action of operators $U\otimes U$.

It turns out there are only two invariant subspaces of $\cH_d\otimes\cH_d$ -
{\it symmetric} and {\it antisymmetric} subspace. A
vector $\psi\in\cH_d\otimes\cH_d$ is called symmetric (antisymmetric)
if $S\psi=\pm\psi$, respectively, where $S$ is the {\it swap} operator.
Let us denote by $P_{\pm}$ the projectors onto the symmetric
and antisymmetric subspaces, respectively.
Consider an orthonormal basis of $\cH_d$ composed of vectors
$\varphi_1,\dots,\varphi_d$. Defining the vectors
$\psi_{j\pm k}=\frac{1}{\sqrt{2}}
(\varphi_j\otimes\varphi_k\pm\varphi_k\otimes\varphi_j)$ for
$j\neq k$, $\psi_{j+j}=\varphi_j\otimes\varphi_j$ and $\psi_{j-j}=0$
we can write
\be
P_\pm=\sum_{jk} \ket{\psi_{j\pm k}}\bra{\psi_{j\pm k}}\, .
\ee
Let us note that vectors $\psi_{j\pm k}$ ($j,k=1,\dots,d$) are forming
an orthonormal basis of $\cH_d\otimes\cH_d$ and
$S\psi_{j\pm k}=\pm\psi_{j\pm k}$. It follows that the
dimensions of symmetric and antisymmetric subspaces
are $d_\pm=d(d\pm 1)/2$, respectively. As a result we obtain that
\be
\label{eq:twirling_form_2}
\cT[X]=a_+(X)P_++a_-(X)P_-\,
\ee
is the spectral form of $\cT[X]$. In order to
verify that Eq.\eqref{eq:twirling_form_1}
and Eq.\eqref{eq:twirling} define the same mapping, it is sufficient
to verify their actions on elements of arbitrary operator basis.
We shall use an orthonormal operator basis consisting of operators
$E_{j\pm k,m\pm n}=\ket{\psi_{j\pm k}}\bra{\psi_{m\pm n}}$.

According to Eq.\eqref{eq:twirling_form_2} $\tr{Y^\dagger\cT[X]}=0$
for arbitrary operator $Y$ orthogonal to $P_\pm$, i.e. $\tr{Y^\dagger P_\pm}=0$.
This identity holds for both expressions of $\cT$. Consequently,
it is sufficient to verify that the values of
$\Delta=\tr{P_\pm\cT[E_{j\pm k,m\pm n}]}$ coincide
for both expressions of the twirling channel given in
Eq.\eqref{eq:twirling_form_1} and in Eq.\eqref{eq:twirling}.
Direct calculation gives
\be
\nonumber
\Delta&=&
\tr{P_\pm\int_{U(d)}dU\, U\otimes U E_{j\pm k,m\pm n}  U^\dagger\otimes U^\dagger}\\
\nonumber
&=&\tr{E_{j\pm k,m\pm n} \int_{U(d)}dU\, U\otimes U P_\pm  U^\dagger\otimes U^\dagger}\\
\nonumber
&=&\tr{E_{j\pm k,m\pm n}P_\pm}
\ee
and, simultaneuously,
\be
\nonumber
\Delta&=&
\frac{\tr{E_{j\pm k,m\pm n}P_+}}{d_+}\tr{P_\pm P_+}+
\frac{\tr{E_{j\pm k,m\pm n}P_-}}{d_-}\tr{P_\pm P_-}\\
\nonumber &=&
\tr{E_{j\pm k,m\pm n}P_\pm}\, .
\ee
That is, the Eqs.\eqref{eq:twirling} and \eqref{eq:twirling_form_1}
determine the same channel.

\end{document}